\documentclass[aps,reprint,pra,longbibliography,groupedaddress]{revtex4-1}
\usepackage{}
\usepackage{amssymb}
\usepackage{amsmath}
\usepackage{dcolumn}
\usepackage{bm}
\usepackage{graphicx}
\usepackage{mathrsfs}
\usepackage[colorlinks,linkcolor=blue,anchorcolor=blue,citecolor=blue,urlcolor=blue]{hyperref}
\usepackage{color}

\begin{document}

\preprint{APS/123-QED}

\title{Dissipation-assisted preparation of steady spin-squeezed states of SiV centers}

\author{Jia-Qiang Chen}
\affiliation{Shaanxi Province Key Laboratory of Quantum Information and Quantum Optoelectronic Devices,
Department of Applied Physics, Xi'an Jiaotong University, Xi'an 710049, China}
\author{Yi-Fan Qiao}
\affiliation{Shaanxi Province Key Laboratory of Quantum Information and Quantum Optoelectronic Devices,
Department of Applied Physics, Xi'an Jiaotong University, Xi'an 710049, China}
\author{Xing-Liang Dong}
\affiliation{Shaanxi Province Key Laboratory of Quantum Information and Quantum Optoelectronic Devices,
Department of Applied Physics, Xi'an Jiaotong University, Xi'an 710049, China}
\author{Xin-Lei Hei}
\affiliation{Shaanxi Province Key Laboratory of Quantum Information and Quantum Optoelectronic Devices,
Department of Applied Physics, Xi'an Jiaotong University, Xi'an 710049, China}
\author{Peng-Bo Li}
 \email{lipengbo@mail.xjtu.edu.cn}
\affiliation{Shaanxi Province Key Laboratory of Quantum Information and Quantum Optoelectronic Devices,
Department of Applied Physics, Xi'an Jiaotong University, Xi'an 710049, China}

\date{\today}

\begin{abstract}
We propose an efficient scheme for generating spin-squeezed states at steady state in a spin-mechanical hybrid system, where an ensemble of SiV centers are coupled to a strongly damped nanomechanical resonator.
We show that, there exists a collective steady state in the system, which is exactly formed by the collective spin states plus the zero excitation state of the mechanical mode.
The generation of the steady spin-squeezed state is based on a dissipative quantum dynamical process in which the mechanical dissipation plays a positive role but without destroying the target state.  We demonstrate  that the spin-squeezed steady state can be deterministically prepared via dissipative means, with the optimal spin squeezing up to $4/N$ in the ideal case, where $N$ is the number of spins. This work provides a promising platform for quantum information processing and quantum metrology.

\end{abstract}

\maketitle


\section{introduction}
Spin-squeezed states are  quantum correlated states where the fluctuation of one component of the total spin is suppressed below the standard quantum limit.
It has been proved that spin-squeezed states are closely related to entangled states \cite{MA201189,PhysRevA.68.012101,PhysRevLett.95.120502,PhysRevA.89.032307,PhysRevA.98.052346}, which lie at the heart of quantum information \cite{2001Natur.409...27B,G_hne_2009,RevModPhys.80.517} and quantum metrology \cite{Appel10960,Louchet_Chauvet_2010,PhysRevA.71.012312,PhysRevApplied.6.034005,RevModPhys.92.015004,RevModPhys.89.035002}.
For these reasons, various schemes for generating spin-squeezed states have been proposed, for instance, the schemes based on nondemolition measurements \cite{PhysRevLett.85.1594,PhysRevA.68.035802,PhysRevLett.104.013601,PhysRevLett.110.163602}, interactions with squeezed light \cite{PhysRevLett.79.4782,PhysRevLett.83.1319,PhysRevLett.88.070404}, and nonlinear interactions \cite{1993PhRvA..47.5138K,PhysRevA.40.2417,PhysRevLett.94.023003}.
In particular, spin squeezing using the one-axis twisting model has been realized in experiments \cite{Gross2010}.
Despite the exciting progress, it still faces enormous challenges to generate spin-squeezed states.
Specifically, due to the disturbance of the environment, most of these spin-squeezed states are temporary and fragile.
Recent works on the generation of target states by ``quantum reservoir engineering" \cite{Verstraete2009,PhysRevA.78.042307,Krauter_2011,PhysRevLett.110.120402,PhysRevA.100.032302} provide us a new avenue for investigating spin squeezing. By appropriately designing the interactions with the reservoir, the dissipation can play a positive role in steering system into the target state.

Solid defects in diamond including nitrogen-vacancy (NV) \cite{PhysRevLett.117.015502,PhysRevApplied.4.044003,PhysRevApplied.5.034010,DOHERTY20131,PhysRevB.85.205203,Song:17,PhysRevA.85.042306,PhysRevX.6.041060,Maze_2011,PhysRevA.96.032342,PhysRevLett.125.153602,PhysRevApplied.10.024011}, silicon-vacancy (SiV) \cite{PhysRevLett.112.036405,PhysRevB.89.235101,Becker2016,PhysRevLett.113.113602,Jahnke_2015,Zhou2017,PhysRevLett.122.063601,PhysRevLett.113.263602,PhysRevResearch.2.013121,Bradac2019,Lemonde_2019}, and germanium-vacancy (GeV) \cite{PhysRevLett.118.223603} color centers have been extensively studied in recent years.
They consist of a substitutional impurity atom replacing the carbon atom and adjacent to vacancies.
The color center exhibits excellent optical properties, which makes it easy to manipulate and detect \cite{PhysRevLett.101.117601,Buckley1212,PhysRevLett.113.263601,PhysRevLett.120.053603}.
More importantly, it has long coherence time \cite{Balasubramanian2009,PhysRevLett.119.223602,Maurer1283,Bar-Gill2013}, which can be used as an excellent qubit and a weak signal detector \cite{PhysRevApplied.11.044026,Taylor2008,Thiel973}.
In addition, the color center embedded in a solid-state device is sensitive to the crystal strain, thus giving rise to a coupling with the mechanical mode.
Hybrid systems directly mediated by crystal strain do not require extra structures, thus avoiding the emergence of extra decoherence.
Single-crystal diamond, with desirable optical and mechanical properties, is a natural carrier of color centers.
Spin-mechanical hybrid quantum systems based on color centers in diamond have been used to produce spin-squeezed states \cite{PhysRevLett.110.156402,PhysRevB.94.214115,PhysRevA.101.042313,doi:10.1002/qute.202000034}.
Although mechanical resonators can achieve high quality factors \cite{doi:10.1063/1.4760274,doi:10.1021/nl302541e,Tao2014} with the great progress of nanofabrication techniques, the produced quantum state is still inevitably spoiled by dissipation and decoherence. Therefore,
it is highly desirable to prepare persistent and high-quality spin-squeezed states. Note that dissipation-assisted schemes for spin squeezing  have been investigated in  cavity QED  \cite{PhysRevLett.110.120402} and photonic waveguide systems \cite{Song:17}.
However, the spin-phononic hybrid system we considered  is fundamentally different from the photonic system.
The solid-state phononic device could have important applications in all-phonon quantum information processing and quantum metrology.

In this work, we propose an efficient scheme to generate spin-squeezed states at steady state in a hybrid system, where an ensemble of SiV centers are coupled to a nanomechanical resonator by crystal strain.
Because of the excellent optical and strain response of the ground states of the SiV center, it is possible  to design two pairs of Raman processes.
In the case of large detunings, we can adiabatically eliminate the high energy levels, thus obtaining the desired interactions.
We first show that the system possesses  a steady squeezed  state with no phonon excitations.
Under the condition of large dissipations for the mechanical resonator, the mechanical mode will continuously absorb energy from the SiV centers and subsequently decay back to its ground state, steering the spins into a steady squeezed state.
This process is verified by numerical simulations.
It is important to note that this process does not depend on the initial state of the system.
Consider the more realistic case where there exists spin dephasing of SiV centers, and the system gives rises to suppressed spin squeezing.
Then, we consider the case of a large number of spins via analytic methods.
Ideally, by choosing the appropriate parameters, the spin squeezing can be up to the magnitude of $4/N$.
In the presence of spin dephasing, the squeezing is suppressed.
When the particle number $N$ increases, the effect of spin dephasing will be reduced so that the spin squeezing is close to the ideal case.
This work presents a useful proposal for generating spin-squeezed states via dissipative
means, which may bring abundant applications in quantum information processing and quantum sensing.

\section{The Model}
\begin{figure}
\includegraphics[width=8cm]{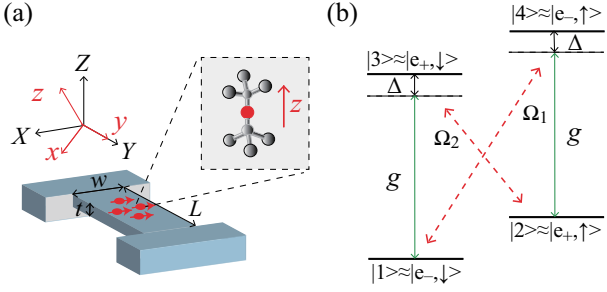}
\caption{\label{fig:1}(Color online)
(a) A diamond nanomechanical resonator with an ensemble of embedded SiV centers.
The length, width and thickness of the beam are $L$, $w$, and $t$, respectively.
Local perpendicular  strain induced by the bending of the beam couples the SiV centers to the mechanical oscillator.
Two sets of coordinates: $(x,y,z)$ corresponds to the internal basis of the SiV center and $(X,Y,Z)$ corresponds to the axes of the beam.
(b) The level structure of the ground state of a SiV center.
Two time-dependent microwave driving fields induce the Raman processes between $|3\rangle\leftrightarrow|2\rangle$, and $|4\rangle\leftrightarrow|1\rangle$. }
\end{figure}

As illustrated in Fig.~\ref{fig:1}(a), the setup under consideration is realized by a doubly clamped nanobeam with an ensemble of embedded SiV centers.
Here the diamond samples used in this study have a $[001]$-oriented top surface, and the beam is along the $[110]$ direction ($Y$ axis).
Thus, there are four possible orientations of SiVs - $[111]$,$[\bar{1}\bar{1}1]$,$[1\bar{1}1]$,$[\bar{1}11]$ - in a diamond crystal.
We choose the SiVs in $z\|[1\bar{1}1]$ direction, and the total Hamiltonian is given by
\begin{eqnarray}\label{ME1}
H=H_{\text{SiV}}+H_{\text{r}}+H_{\text{I}}+H_{\text{d}},
\end{eqnarray}
where the individual terms represent the Hamiltonian of the SiV centers, the mechanical mode of the diamond beam, the strain coupling of the SiV  centers to the mechanical mode, and the classical driving, respectively.

For the SiV center, the electronic ground states consist of four spin-orbit degenerate states $|e_{x},\downarrow\rangle,|e_{x},\uparrow\rangle,|e_{y},\downarrow\rangle$ and $|e_{y},\uparrow\rangle$, with spin $S=1/2$.
In the presence of an external magnetic field $\vec{B}=B\vec{e}_{z}$, the level structure is affected by the spin-orbit coupling (SO), the Jahn-Teller (JT) effect and Zeeman effect. Thus the Hamiltonian of the SiV center is given by
\begin{eqnarray}\label{ME2}
H_{\text{SiV}}=-\lambda_{\text{SO}}\vec{L}\cdot\vec{S}+H_{\text{JT}}+f\gamma_{L}BL_{z}+\gamma_{S}BS_{z}.
\end{eqnarray}
Here, $\gamma_L$ and $\gamma_S$ are the orbital and
spin gyromagnetic ratios, respectively. The spin-orbit coupling with the strength $\lambda_{\text{SO}}/2\pi\simeq45\text{GHz}$ \cite{PhysRevLett.120.213603} is much stronger than the other terms.
Therefore, we first consider the spin-orbit coupling and treat the JT interaction and Zeeman interaction as disturbance.
In the above basis, as the projections of angular momentum $L_{x},L_{y}$ are vanished, the first term is reduced to $-\lambda_{\text{SO}} L_{z}S_{z}$.
The corresponding eigenstates are two doublet states ${|e_{-},\downarrow\rangle,|e_{+},\uparrow\rangle}$ and ${|e_{+},\downarrow\rangle,|e_{-},\uparrow\rangle}$, where $|e_{\pm}\rangle=(|e_{x}\rangle\pm i|e_{y}\rangle)/\sqrt{2}$ are the eigenstates of the orbital angular momentum operator $L_{z}$.
Due to the JT interaction strength $\Upsilon\ll\lambda_{\text{SO}}$, we neglect the mixing of energy levels caused by the JT effect and only consider the change of eigenenergy $\omega=\sqrt{\lambda_{\text{SO}}^{2}+4\Upsilon^{2}}$.
As the orbital Zeeman interaction is suppressed by $f\approx0.1$, it can be neglected and the Zeeman effect simply splits the spin states with $\omega_{B}=\gamma_{S}B$.
Thus, the Hamiltonian for a single SiV center reads \cite{PhysRevLett.120.213603,PhysRevB.94.214115}
\begin{eqnarray}\label{ME3}
H_{\text{SiV}}=\omega_{B}|2\rangle\langle2|+\omega|3\rangle\langle3|+(\omega+\omega_{B})|4\rangle\langle4|.
\end{eqnarray}
Besides, we add two controlled fields $\Omega_{1/2}$ with frequencies $\omega_{1/2}$ to drive the transitions $|1\rangle,|2\rangle\rightarrow|4\rangle,|3\rangle$, respectively.
So, the corresponding Hamiltonian is given by
\begin{eqnarray}\label{ME4}
H_{\text{d}}=\Omega_{1}|4\rangle\langle1|e^{-i\omega_{1}t}+\Omega_{2}|3\rangle\langle2|e^{-i\omega_{2}t}+\text{H.c.}.
\end{eqnarray}

For the mechanical resonator, we choose a doubly clamped diamond beam with a length $L$, width $w$, and thickness $t$, satisfying $L\gg w,t$.
In the case of small beam displacements, the strain $\epsilon=\epsilon_{0}(a+a^{+})$ is linear, where $\epsilon_{0}$ is strain induced by the zero-point motion of the beam, and $a^{+}(a)$ is the creation (annihilation) operator for the phonon mode with frequency $\omega_{0}$.
The corresponding Hamiltonian is
\begin{eqnarray}\label{ME5}
H_{\text{r}}=\omega_{0}a^{+}a.
\end{eqnarray}
The strain results in the mixture of the orbital states \cite{PhysRevLett.120.213603,PhysRevB.97.205444} (see Appendix A).
In the rotating-wave approximation, the Hamiltonian for the strain coupling reads
\begin{eqnarray}\label{ME6}
H_{\text{I}}=g(a^{+}J_{-}+aJ_{+}),
\end{eqnarray}
where $J_{-}=(J_{+})^{\dag}=|1\rangle\langle3|+|2\rangle\langle4|$ is the spin-conserving lowering operator.
Finally, combine the Hamiltonian of SiV and mechanical mode into free Hamiltonian $H_{0}=H_{\text{SiV}}+H_{\text{r}}$.

The key idea of this scheme is to design an appropriate interaction and bring the system into the desired state by means of dissipations.
We first consider the  case where there is only mechanical dissipation.
Assuming that the characteristic time scale of
the system Hamiltonian is much longer than the reservoir correlation time, the environment can therefore be considered as a memoryless reservoir. So we can get the Markovian master equation
\begin{eqnarray}\label{ME7}
\frac{d\rho}{dt}=-i[H,\rho]+\kappa \mathcal{D}[a]\rho,
\end{eqnarray}
where $\mathcal{D}[o]\rho=2o\rho o^{\dag}-\rho o^{\dag}o-o^{\dag}o\rho$ is the standard Lindblad operator and $\kappa$ is the mechanical dissipation rate.

To get more insight into the underlying physics, we first apply several transformations to the Hamiltonian to get an effective Hamiltonian.
First, in a rotating frame where the classical fields are not oscillating in time just by the following unitary operation $U_{R}=e^{iRt}$, with
\begin{eqnarray}\label{ME8}
R=\omega_{1}|4\rangle\langle4|+\omega_{2}|3\rangle\langle3|,
\end{eqnarray}
the  Hamiltonian for a single SiV center becomes $H^{R}=U_{R}HU_{R}^{\dag}+i\dot{U}_{R}U_{R}^{\dag}$.

Next, in the case of large detunings $\Delta\gg\Omega_{1/2}$, we apple the Schrieffer-Wolff (SW) transformation \cite{PhysRev.149.491} $U_{S}=e^{S}$ with $S=\frac{\Omega_{1}}{\Delta}(|4\rangle\langle1|-|1\rangle\langle4|)+\frac{\Omega_{2}}{\Delta}(|3\rangle\langle2|-|2\rangle\langle3|)$ and $\Delta=\omega+\Delta_{B}-\omega_{1}=\omega-\Delta_{B}-\omega_{2}$, which satisfy $[H_{0}^{R},S]=H_{\text{d}}^{R}$. Keeping the terms up to $(\frac{\Omega_{1/2}}{\Delta})^2$ and ignoring the energy shifts, we obtain
\begin{eqnarray}\label{ME9}
H^{S}&=&H_{0}^{R}+\frac{1}{2}[H_{\text{d}}^{R},S]+e^{S}H_{\text{I}}^{R}e^{-S}.
\end{eqnarray}
Finally, turning into the interaction picture by performing the unitary transformation $U_{I}=e^{-iH_{0}^{R}t}$, we have the following Hamiltonian
\begin{eqnarray}\label{ME10}
H^{I}&=&ga^{+}(|1\rangle\langle3|e^{-i\Delta t}+\frac{\Omega_{1}}{\Delta}|4\rangle\langle3|e^{i2\omega_{B}t}-\frac{\Omega_{2}}{\Delta}|1\rangle\langle2|) \notag\\
&+&ga^{+}(|2\rangle\langle4|e^{-i\Delta t}+\frac{\Omega_{2}}{\Delta}|3\rangle\langle4|e^{-i2\omega_{B}t}-\frac{\Omega_{1}}{\Delta}|2\rangle\langle1|) \notag\\
&+&\text{H.c.}.
\end{eqnarray}

Under the condition $\Delta,\omega_B\gg g,\Omega_{1/2}$, the fast oscillating terms $e^{\pm i\Delta t}$ and $e^{\pm i2\omega_B t}$ can be completely discarded in the rotating-wave approximation.
In this case, we implement  the standard Raman-transition process with the $\Lambda$-type and can eliminate the high energy levels under the large detuning condition.
Therefore, the states $|3\rangle, |4\rangle$ can be safely eliminated to obtain the effective Hamiltonian
\begin{eqnarray}\label{ME11}
H_{\text{eff}}=\frac{g\Omega}{\Delta}a^{+}(\sin\theta\sigma_{+}+\cos\theta\sigma_{-})+\text{H.c.},
\end{eqnarray}
where $\sin\theta=\Omega_{1}/\Omega$ with $\Omega^{2}=\Omega_{1}^{2}+\Omega_{2}^{2}$ and $\sigma_{-}=(\sigma_{+})^{\dag}=|1\rangle\langle2|$. We further introduce the collective operators $S_{-}=(S_{+})^{\dag}=\sum_{j}\sigma_{-}^{j}$ and $S_{z}=\frac{1}{2}\sum_{j}(|2\rangle_{j}\langle2|-|1\rangle_{j}\langle1|)$, which satisfy the usual angular momentum commutation relations
\begin{eqnarray}\label{ME12}
[S_{+},S_{-}]=2S_{z},\,[S_{z},S_{\pm}]=\pm S_{\pm}.
\end{eqnarray}
For an ensemble of SiV centers, the sum runs over all SiV centers, and we obtain the total effective Hamiltonian
\begin{eqnarray}\label{ME13}
H_{\text{eff}}&=&\frac{g\Omega}{\Delta}a^{+}(\sin\theta S_{+}+\cos\theta S_{-})+\text{H.c.} \notag\\
&=&\frac{g\Omega}{\Delta}a^{+}D_{-}+\text{H.c.},
\end{eqnarray}
where $D_{-}=D_{+}^{\dag}=\sin\theta S_{+}+\cos\theta S_{-}$.
The coupling $g\Omega/\Delta$ is therefore an effective Rabi frequency.
Meanwhile, the effective master equation is given by
\begin{eqnarray}\label{ME14}
\frac{d\rho}{dt}=-i[H_{\text{eff}},\rho]+\kappa \mathcal{D}[a]\rho.
\end{eqnarray}
The Hamiltonian $H_{\text{eff}}$ describes
 the transition between the states $\{|1\rangle,|2\rangle\}$, creating or annihilating a phonon, while the Lindblad term will drive the transition from the phonon state $|n\rangle$ to $|n-1\rangle$.
Hence, the combined effect of the unitary and dissipative
dynamics drives
the system to the state $|\Psi\rangle=|\psi\rangle|0\rangle$, i.e., the tensor product of the steady state $|\psi\rangle$ of SiV centers and the phonon vacuum state $|0\rangle$.
Since the steady state satisfies $d\rho/dt=0$, the steady state of SiV centers should obey the equation
\begin{eqnarray}\label{ME15}
D_{-}|\psi\rangle=0.
\end{eqnarray}

Assuming the total spin of the system $S=N/2$, one can expand the steady state in the basis of the eigenstates of $S_{z}$ as $|\psi\rangle=\sum_{m}c_{m}|S=N/2,\,S_{z}=-N/2+m\rangle$.
By solving the equation (\ref{ME15}), we get the recursion relation between $c_{m}$ and $c_{m+2}$.
Assuming $c_{1}=0$, we only have the even terms left
\begin{eqnarray}\label{ME16}
c_{m=2n}=(-1)^{n}\tan^{n}\theta\binom{N/2}{n}\binom{N}{2n}^{-1/2}c_{0}.
\end{eqnarray}
Here $\binom{A}{B}=\frac{A!}{B!(A-B)!}$ are the binomial coefficients and the normalization condition of the state $\sum_{m}|c_{m}|^{2}=1$ determines the value of $c_{0}$.
Note that the mean spin of the steady state is along the $-z$ direction, and the fluctuation $(\Delta S_{x}^{2})$ is suppressed.
This is exactly a steady spin-squeezed state that we expect.
To quantify the degree of squeezing, we use the spin squeezing parameter introduced by Wineland et al \cite{PhysRevA.50.67,PhysRevA.46.R6797}.
For this system, the spin squeezing parameter reduces to
\begin{eqnarray}\label{ME17}
\xi^{2}=\frac{N\langle S_{x}^{2}\rangle}{|\langle S_{z}\rangle|^{2}}.
\end{eqnarray}

\section{Numerical simulations}
\begin{figure}[t]
\includegraphics[width=8.5cm]{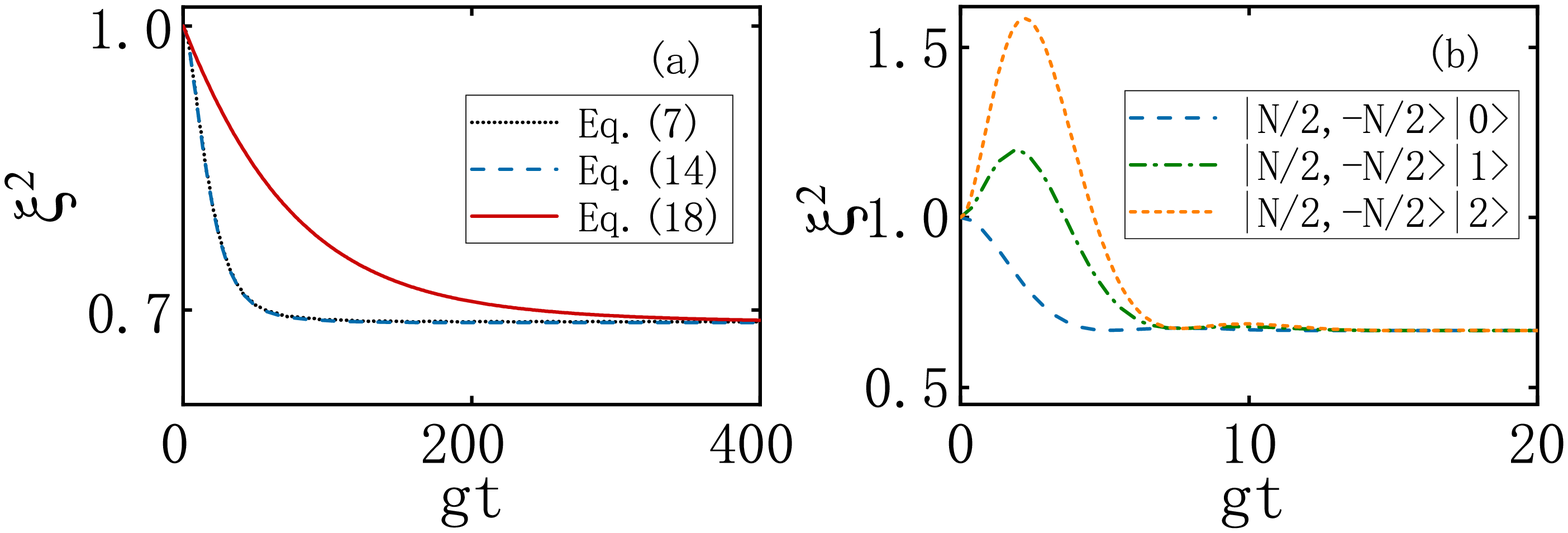}
\caption{\label{fig:2}(Color online)
(a) The spin squeezing parameter calculated from equation~(\ref{ME7}), (\ref{ME14}) and~(\ref{ME18}) with $N=4$ and $\kappa=0.5g$. The initial state is that all the SiV centers are in the state $|1\rangle$ and the mechanical mode is in the vacuum state. (b) The spin squeezing parameter from equation~(\ref{ME14}) for different initial states, where  $N=100$ and $\kappa=g$. The other parameters are $\Delta=20g$, $\Omega=g$ and $\tan\theta=0.2$.}
\end{figure}

If the mechanical dissipation is sufficiently large, the mechanical mode can be adiabatically eliminated.
The reduced master equation for the SiV centers is
\begin{eqnarray}\label{ME18}
\frac{d\rho_{N}}{dt}=\gamma \mathcal{D}[D_{-}]\rho_{N},
\end{eqnarray}
where $\gamma=g^{2}\Omega^{2}/\Delta^{2}\kappa$ is the collective decay rate induced by mechanical dissipation.

Fig.~\ref{fig:2}(a) shows the results given by equation~(\ref{ME7}), (\ref{ME14}) and~(\ref{ME18}).
We can find that the values for the squeezing parameter calculated from the effective Hamiltonian (13)   are almost the same as those obtained from the  Hamiltonian (1),  demonstrating the reasonability of adiabatic eliminations of the high energy levels.
Furthermore, we find that the squeezing parameter at the steady state is the same for the two cases obtained from
equation~(\ref{ME14}) and equation~(\ref{ME18}), respectively.  This demonstrates that, in the large dissipation limit, the phononic degree of freedom can be adiabatically eliminated from the dynamics.
Fig.~\ref{fig:2}(b) shows the results for the squeezing parameter obtained from equation~(\ref{ME14}) with different initial states.
The system almost simultaneously evolves into the unique steady spin-squeezed state independent of the initial state.
Therefore, we can conclude that equation~(\ref{ME18})  approximately equals to equation~(\ref{ME14}) in the large dissipation limit.
Subsequent discussions will be uniformly based on equation~(\ref{ME18}).

\begin{figure}[t]
\includegraphics[width=8cm]{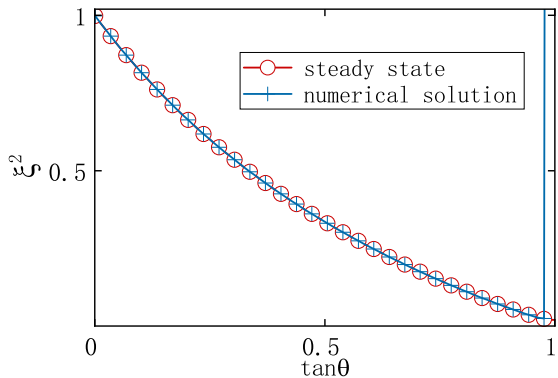}
\caption{\label{fig:3}(Color online)
The spin squeezing parameter as a function of $\tan\theta$ with $N=100$ spins.}
\end{figure}
Assuming the initial state as $|N/2,-N/2\rangle$, we plot the spin squeezing parameter of the numerical solution \cite{Perarnau_Llobet_2020} compared with the steady state $|\psi\rangle$ as a function of $\tan\theta=\Omega_{1}/\Omega_{2}$ in Fig.~\ref{fig:3}.
As expected, the numerical results coincides with the steady state case, indicating
that the system evolves to the desired spin-squeezed
state.

When $\tan\theta=0$, the spin squeezing parameter $\xi^{2}=1$ is the standard quantum limit (SQL) corresponding to a spin coherence state.
When $\tan\theta<1$, the numerical result completely coincides with the steady state case, and the spin squeezing parameter is less than 1, indicating that the steady state is a spin-squeezed state.
The degree of spin squeezing can be enhanced continuously with the increasing of $\tan\theta$.
When $\tan\theta\rightarrow1$, there is a large deviation between these two cases, since the evolution time of the system will increase rapidly and become infinite at $\tan\theta=1$.
This is due to the fact that the evolution rate is proportional to $\cos(2\theta)$, which will be seen in the following analysis.

\begin{figure}[t]
\includegraphics[width=8cm]{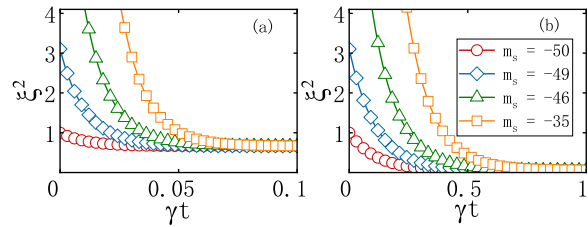}
\caption{\label{fig:4}(Color online)
Time evolution of the squeezing parameter $\xi^{2}$ for initial states with different $m_{s}$.
The parameters are $N=100$, (a) $\tan\theta=0.2$ and (b) $\tan\theta=0.9$.
}
\end{figure}
As the Hamiltonian commutes with the total spin $S^{2}$ and the master equation doesn't break the conservation of total spin, the total spin will be a constant of motion.
Therefore, this steady squeezed state could be produced independent of the $z$ component of the angular momentum.
The time evolution of the spin squeezing parameter for an ensemble of spins with different $S_{z}$ is shown in Fig.~\ref{fig:4}.
When $t=0$, all the spin squeezing parameters $\xi^{2}\geq1$, which means that they are not squeezed at the initial time.
Then under the assistance of mechanical dissipation, the system starting from different initial states  is steered into the same steady squeezed state. It indicates that one total spin corresponds to a common steady state.
Compared to Fig.~\ref{fig:4}(a), Fig.~\ref{fig:4}(b) gives a reduced squeezing parameter, which is exactly the result shown in Fig.~\ref{fig:3}.
In addition, since the effective decay rate decreases as $\tan\theta$ increases, the evolution time increases.

Now we consider the more realistic case, where  the dephasing of SiV centers with a rate $\Gamma$ is taken into account.
The corresponding master equation can be obtained  as
\begin{eqnarray}\label{ME19}
\frac{d\rho_{N}}{dt}=\gamma \mathcal{D}[D_{-}]\rho_{N}+\Gamma \sum_{j}\mathcal{D}[\sigma_{z}^{j}]\rho_{N}.
\end{eqnarray}
Here the Pauli operator $\sigma_{z}=|2\rangle\langle2|-|1\rangle\langle1|$.
Such a single particle process may break the conservation of the total spin.
\begin{figure}[b]
\includegraphics[width=8cm]{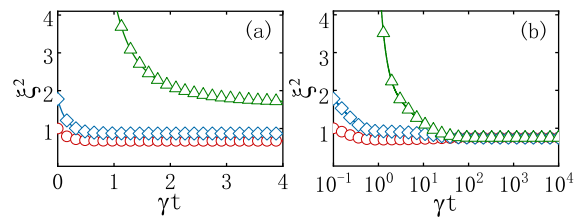}
\caption{\label{fig:5}(Color online)
Time evolution of the squeezing parameter $\xi^{2}$ starting from initial states with different total spins.
Here the parameters are $N=8$, $\tan\theta=0.2$.
(a) $\Gamma=0$. The steady state does depend on the total spins of the initial state.
(b) $\Gamma=0.1\gamma$.
The spin dephasing breaks the dependence on the total spin. }
\end{figure}
We present the numerical solutions of equation ~(\ref{ME19}) with $N=8$ in Fig.~\ref{fig:5}.
Fig.~\ref{fig:5}(a) shows the time evolution of the squeezing parameter starting from initial states with a different
total spin in the case of no dephasing, i.e., $\Gamma=0$.
We find that all the curves decrease continuously with time, but they correspond to a different final state.
The bottom curve, corresponding to the case of the total spin $S=N/2$, gives the spin squeezing parameter $\xi^{2}\approx0.67$.

In Fig.~\ref{fig:5}(b), the spin dephasing with the rate $\Gamma=0.1\gamma$ is taken into account.
With the same initial conditions as in Fig.~\ref{fig:5}(a), the spin squeezing parameters eventually evolve to the same value, which implies that spin dephasing couples different total spin-S manifolds and the system has a unique steady state independent of the total spins.
Compared to Fig.~\ref{fig:5}(a), the first part of the curves are remarkably similar, where the collective decay induced by phonon dominates the entire evolution.
As time goes on, although the dephasing rate is very small, it still shows a great influence: a steady spin-squeezed state completely independent of the initial state is obtained.
If we focus on the case with the total spin $S=N/2$, the existence of spin dephasing reduces the degree of squeezing and seems to compete with the squeezing process caused by the phonon decay.
But the dependence of the system on the initial state is completely eliminated.
In addition, when using a spin ensemble to prepare the spin squeezing, the degree of squeezing usually increases with the number of particles.

\section{approximate analytic solutions}
In order to measure the  degree of squeezing in the case of large particle numbers, it is better to find the expression for the spin squeezing parameter under proper approximations.
Note that the spin squeezing parameter $\xi^{2}$ is given by the ratio of $\langle S_{x}^{2}\rangle$ to $\langle S_{z}\rangle^{2}$, and we first neglect spin dephasing and solve for the time evolution of $\langle S_{x}^{2}\rangle$ and $\langle S_{z}\rangle$ from the equation ~(\ref{ME18})
\begin{eqnarray}\label{ME20}
\frac{d\langle S_{z}\rangle}{dt}=-2\gamma(\cos^{2}\theta\langle S_{+}S_{-}\rangle-\sin^{2}\theta\langle S_{-}S_{+}\rangle)
\end{eqnarray}
\begin{eqnarray}\label{ME21}
\frac{d\langle S_{x}^{2}\rangle}{dt}&=&-2\gamma(\sin\theta-\cos\theta)\notag\\
&\times&\langle(\sin\theta S_{-}+\cos\theta S_{+})(S_{x}S_{z}+S_{z}S_{x})+\text{H.c.}\rangle.
\notag\\
\end{eqnarray}
\begin{figure}[b]
\includegraphics[width=8cm]{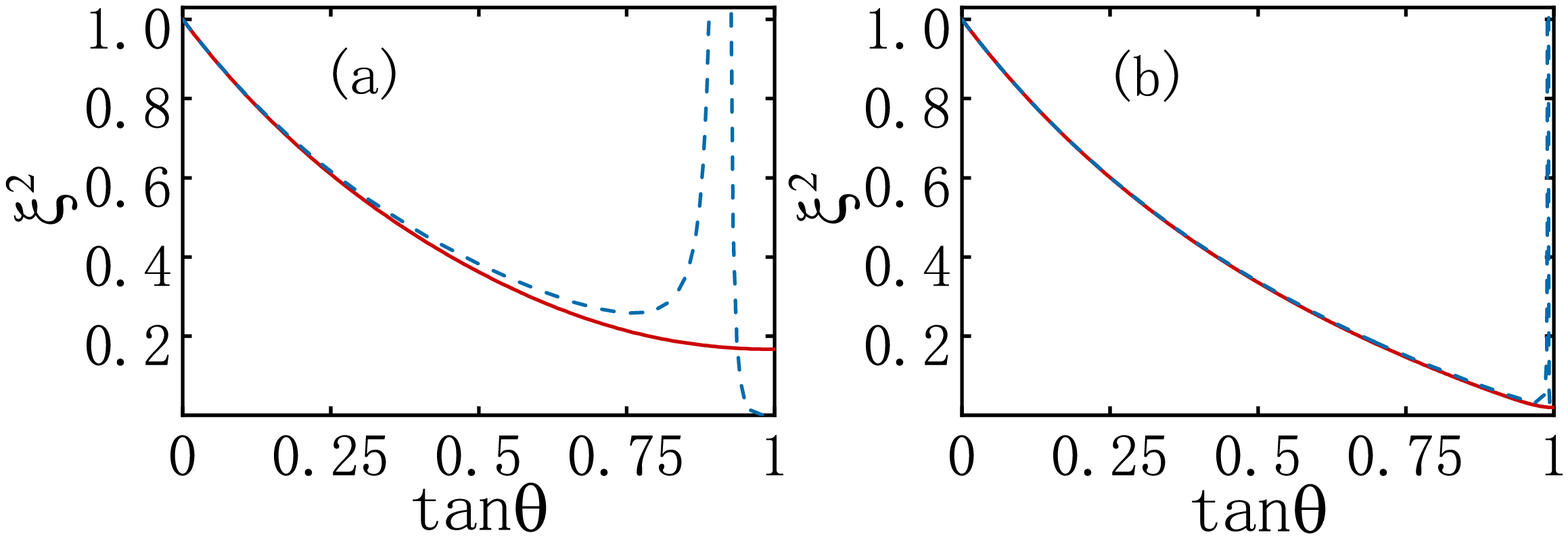}
\caption{\label{fig:6}(Color online)
Spin squeezing $\xi^{2}$ for (a) $N=10$ and (b) $N=100$ spins.
Exact solution of steady state (solid red line) and the analytic approximate solution (dashed blue line). }
\end{figure}

Due to the main spin along $-z$ axis, we use linearization $\langle S_{z}\rangle\approx-N/2$ and define the small fluctuation as $\delta S_{z}=S_{z}+N/2$ to get the simplified equations
\begin{eqnarray}\label{ME22}
\frac{d\langle\delta S_{z}\rangle}{dt}=-2N\gamma\cos(2\theta)[\langle\delta S_{z}\rangle-\frac{\sin^{2}\theta}{\cos(2\theta)}]
\end{eqnarray}
\begin{eqnarray}\label{ME23}
\frac{d\langle S_{x}^{2}\rangle}{dt}=-2N\gamma\cos(2\theta)\{\langle S_{x}^{2}\rangle-\frac{N[1-\sin(2\theta)]}{4\cos(2\theta)}\}.
\end{eqnarray}
We can see that they decay exponentially in time, and $\gamma_{eff}=2N\gamma\cos(2\theta)$ is the effective rate.
When the system reaches steady state, the expectation values of the two operators will not change.
We can substitute the steady state solution of the equations into equation~(\ref{ME17})
\begin{eqnarray}\label{ME24}
\xi^{2}=\frac{N^{2}[1-\sin(2\theta)]\cos(2\theta)}{[N\cos(2\theta)-2\sin^{2}\theta]^{2}}.
\end{eqnarray}

In Fig.~\ref{fig:6}, we compare this approximate expression with the exact solution of the steady state $|\psi\rangle$.
The approximation is valid only if $\tan\theta$ is small.
When $\tan\theta$ gets large, the approximate result gradually deviates and even breaks the Heisenberg limit at $\tan\theta=1$.
The deviation comes from two reasons: (1) with the increase of $\tan\theta$, the probability of occupying the highly excited states increases,  as shown in the numerical solution.
That means $\langle S_{z}\rangle\rightarrow0$, resulting in a rapid increase of spin squeezing parameter.
(2) at $\tan\theta\sim1$, the approximate result gives rise to $\langle S_{z}\rangle=\frac{\sin^{2}\theta}{\cos(2\theta)}-\frac{N}{2}\rightarrow\infty$, an infinite denominator, which drives the spin squeezing parameter down to zero.

\begin{figure}[t]
\includegraphics[width=8cm]{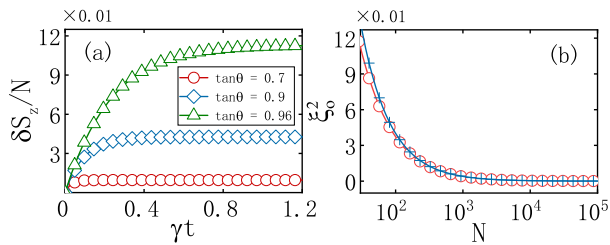}
\caption{\label{fig:7}(Color online)
(a) Time evolution of $\delta S_{z}/N$ for different $\tan\theta$ with $N=100$.
(b) The optimal squeezing $\xi^2_o$ as a function of $N$.
The red curve corresponds to the analytic approximate solution with $\tan^{2}\theta=\frac{N}{N+10}$.
The blue curve  corresponds to $4/N$.
}
\end{figure}

We plot the time evolution of $\langle\delta S_{z}\rangle/N$ with different $\tan\theta$ in Fig.~\ref{fig:7}(a) to obtain the region where the approximation is reasonable and desirable.
The linearization that we used is equivalent to $\langle\delta S_{z}\rangle/N\approx0$.
As $\tan\theta$ increases,  $\langle\delta S_{z}\rangle/N$ at the steady state increases so rapidly that the approximation completely fails.
Therefore, here we choose $\langle\delta S_{z}\rangle/N\leq0.1$ to satisfy the linear approximation.
One can get $\tan^{2}\theta\leq\frac{N}{N+10}$ and verify  that the approximate solution is valid in Fig.~\ref{fig:6}.
Although this condition is approximate, it can be verified that $\tan^{2}\theta=\frac{N}{N+10}$ is close to the region where the minimum value occurs. Therefore, we assume that the  spin squeezing  parameter near this region  is the optimal parameter that leads to maximal squeezing.
The corresponding squeezing parameter $\xi^2_o$ (red line) is shown in Fig.~\ref{fig:7}(b).
As $N$ increases, the spin squeezing parameter decreases dramatically and it can be found that it fits well with $\xi^{2}_o=4/N$ (blue line) for very  large $N$.
\begin{figure}[b]
\includegraphics[width=8cm]{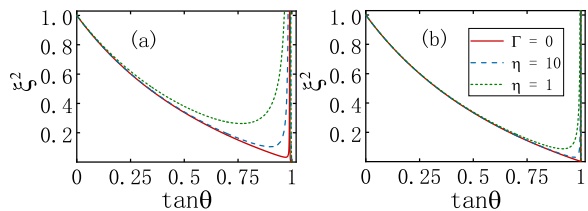}
\caption{\label{fig:8}(Color online)
The analytic approximate solution  as a function of $\tan\theta$ in the presence of dephasing for (a) $N=100$ and (b) $N=1000$ spins.
}
\end{figure}

Finally, we take into account the spin dephasing to study the degree of squeezing in the case of very large $N$.
Once again, by using equation~(\ref{ME19}) we obtain
\begin{eqnarray}\label{ME25}
\langle\delta S_{z}\rangle_{\infty}=\frac{\sin^{2}\theta}{\cos(2\theta)},
\end{eqnarray}
\begin{eqnarray}\label{ME26}
\langle S_{x}^{2}\rangle_{\infty}=\frac{N}{4}\frac{N\eta[1-\sin(2\theta)]+4}{N\eta\cos(2\theta)+4},
\end{eqnarray}
where $\eta=\gamma/\Gamma$ is the single spin cooperativity.
Note that the spin squeezing parameter should be equal to equation~(\ref{ME24}) when $N\eta$ is large.
We present the calculation of the spin squeezing parameter with different $\eta$ in Fig.~\ref{fig:8}.
We find that the dephasing only slightly affects the spin squeezing parameter for a wide parameter range.
It competes with the squeezing process induced by the mechanical dissipation. However, as the value of $\tan\theta$ increases, the
dephasing effect will be significant.
The spin squeezing parameter has a minimum value corresponding to the optimal squeezing in the presence of dephasing.
By increasing the number of SiV centers and decreasing the dephasing rate, the system will work very well even for large $\tan\theta$.

\section{Experimental feasibility of this scheme}
In order to examine the feasibility of this scheme in  experiment, we now discuss the relevant parameters.
For the beam, we consider a single-crystal diamond nanomechanical resonator with dimensions $(L,w,t)=(6.29,0.5,0.5)\,\mu \text{m}$. The material properties are Young's modulus $E\approx1.05\,\text{TPa}$, mass density $\rho\approx3500\,\text{kg}\cdot\text{m}^{-3}$ and Poisson ratio $\nu=0.2$. As the splitting of the orbital states of SiV centers is about $46$ GHz, the corresponding resonance frequency of the bending mode is $\omega_0 /2\pi\approx45.9$ GHz. Generally, the SiV centers in a diamond crystal have four different orientations. In the case of static magnetic fields along the $[1\bar{1}1]$- orientation, these four types of SiV centers have different Zeeman splittings. Only the SiV centers in  $[1\bar{1}1]$- orientation resonate with the driving field and are considered while other SiV centers can be neglected due to off-resonance. In this case, the coupling between a single SiV center and the mechanical mode is given by $g=(1+\nu)d\epsilon_0\approx2\pi\times10$ MHz.

In addition, we assume the driving field $\Omega_1/2\pi=6.7$ MHz, $\Omega_2/2\pi=7.4$ MHz, i.e., $\Omega/2\pi=10$ MHz, and $\tan\theta=0.9$. And the detuning $\Delta/2\pi$ in our scheme is about 200 MHz, thus satisfying the condition $\Delta\gg g,\,\Omega$. The mechanical dissipation rate is $\kappa/2\pi\sim10$ MHz. For single SiV center, the effective decay rate induced by the mechanical dissipation is $\gamma=g^2\Omega^2/\Delta^2\kappa\sim2\pi\times25$ kHz.
We assume the temperature $T=100$ mK, and thus the average phonon number is $n\approx0$, which leads to the correlation time of thermal reservoir $t_R\approx\hbar/k_BT$. Since $t_S/t_R\sim10^3$, the Markovian approximation is justified.

At 100 mK temperature, the spin dephasing rate of single SiV centers is $\Gamma/2\pi\sim100$ Hz, which means the spin coherence time $T_2\sim10$ ms. According to Fig.~\ref{fig:4}, the time for the system to reach the steady spin-squeezed state is about 40 $\mu$s, which is much shorter than the spin coherence time. Experimentally, the SiV center density has reached 8 ppm \cite{doi:10.1063/1.4967189}, so it is easy to reach $10^4$ color centers in our scheme. In this case, we predict that the spin squeezing parameter can be up to $4\times10^{-4}$ with the appropriate parameters.

\section{Conclusion}
In conclusion, we have proposed a scheme to generate steady spin-squeezed state in a spin-mechanical hybrid system.
We find a collective steady state in the system, which is exactly formed by the collective spin states plus the zero excitation state of the mechanical mode. In the case of large detuning, the system of total spin $S=N/2$ can be steered to a steady squeezed state with the assistance of mechanical dissipation.
In the ideal case, the squeezing $\xi_o^2\sim4/N$ can be achieved.
In the presence of spin dephasing, the spin squeezing is reduced.
The generation of the steady spin-squeezed state is not dependent on the initial state of the system and the degree of spin squeezing can reach the result in the ideal case as $N$ increases. This spin-mechanical system provides a promising platform for studying the nonclassical state.

\section*{Acknowledgments}

This work is supported by the National Natural Science Foundation of
China under Grant No.
11774285, and Natural Science Basic Research Program of Shaanxi (Program No.
2020JC-02).

\begin{appendix}

\section{Strain coupling}
The detailed derivation for the strain coupling between SiV centers and mechanical resonators has been discussed in Refs \cite{PhysRevLett.120.213603,Ovartchaiyapong2014,PhysRevB.97.205444}.
Here we follow the discussions and present the key results.
In what follows, we use two sets of coordinates: the internal basis of the SiV center and the axes of the diamond crystal ($X\|[1\bar{1}0], Y\|[110], Z\|[001]$).
There are four possible directions of the color center inlaid on the diamond sample, $[\bar{1}11]$, $[1\bar{1}1]$, $[\bar{1}\bar{1}1]$ and $[111]$ respectively.
For a long and thin beam, the bending modes can be described by elasticity theory and Euler-Bernoulli flexure theory.
The wave equation for beam deflections of the neutral axis obeys
\begin{eqnarray}\label{MA1}
EI\frac{\partial^{4}U}{\partial Y^{4}}=-\rho A\frac{\partial^{2}U}{\partial t^{2}},
\end{eqnarray}
where $U$ represents the beam deflection in the $Z$ direction and $Y$ is along the length of the beam. Here, $E\approx1.05\,\text{TPa}$ is Young's modulus of diamond, $I=wt^{3}/12$ is the moment of inertia of the beam, $\rho\approx3500\,\text{kg}\cdot \text{m}^{-3}$ is the mass density of diamond, and $A=wt$ is the cross sectional area in the transverse plane. Solutions of the above equation for a doubly clamped beam take the form $U_{n}(t,Y)=u_{n}(Y)e^{-i\omega_{n}t}$.
According to the boundary conditions $u(0)=u^{'}(0)=u(L)=u^{'}(L)=0$, $u_{n}(Y)$ is given by
\begin{eqnarray}\label{MA2}
u_{n}(Y)&=&\cos k_{n}Y-\cosh k_{n}Y-\frac{\cos k_{n}L-\cosh k_{n}L}{\sin k_{n}L-\sinh k_{n}L} \notag\\
&\times&(\sin k_{n}Y-\sinh k_{n}Y).
\end{eqnarray}
The wavenumbers $k_{n}$ satisfy $\cos k_{n}L\cosh k_{n}L=1$, and the eigenfrequencies are given by
\begin{eqnarray}\label{MA3}
\omega_{n}=k_{n}^{2}\sqrt{\frac{EI}{\rho A}}.
\end{eqnarray}
We consider the frequency near $46\,\text{GHz}$, which is obtained when $kL=67.55$.
For small beam displacements, the strain induced by the beam motion is linear $\epsilon=\epsilon_{0}(a+a^{+})$. Here, $\epsilon_{0}$ is the strain induced by the beam zero-point motion and $a$ is the annihilation operator of the phonon mode.
Assuming that the SiV center is located in the midpoint along the beam and away from the neutral axis of the beam, i.e., $Y=L/2,R_{0}\approx t/2$, the strain induced by the zero point motion of the beam is
\begin{eqnarray}\label{MA4}
\epsilon_{0}=-R_{0}\sqrt{\frac{\hbar}{2\rho AL\omega_{n}}}\frac{\partial^{2}u_{n}(L/2)}{\partial Y^{2}}\approx8\times10^{-9}.
\end{eqnarray}
In the beam's basis, the strain tensor is given by
\begin{eqnarray}\label{MA5}
\epsilon_{b}=\begin{bmatrix} -\nu\epsilon&0&0 \\ 0&\epsilon&0 \\ 0&0&-\nu\epsilon \end {bmatrix},
\end{eqnarray}
where $\nu=0.2$ is the Poisson ratio.
Now we need to transform the above strain tensor from the beam's basis into the SiV's basis. For the SiVs oriented along the $[111]$ and $[\bar{1}\bar{1}1]$ direction, the transformed strain tensors are
\begin{eqnarray}\label{MA6}
\epsilon_{[111]}=\begin{bmatrix} \frac{1}{3}(1-2\nu)\epsilon&0&-\frac{\sqrt{2}}{3}(1+\nu)\epsilon \\ 0&-\nu\epsilon&0 \\ -\frac{\sqrt{2}}{3}(1+\nu)\epsilon&0&\frac{1}{3}(2-\nu)\epsilon \end {bmatrix}
\end{eqnarray}
\begin{eqnarray}\label{MA7}
\epsilon_{[\bar{1}\bar{1}1]}=\begin{bmatrix} \frac{1}{3}(1-2\nu)\epsilon&0&\frac{\sqrt{2}}{3}(1+\nu)\epsilon \\ 0&-\nu\epsilon&0 \\ \frac{\sqrt{2}}{3}(1+\nu)\epsilon&0&\frac{1}{3}(2-\nu)\epsilon \end {bmatrix}.
\end{eqnarray}
For the SiVs oriented along the $[\bar{1}11]$ and $[1\bar{1}1]$ directions, the transformed strain tensors are the same
\begin{eqnarray}\label{MA8}
\epsilon_{t}=\begin{bmatrix} -\nu\epsilon&0&0 \\ 0&\epsilon&0 \\ 0&0&-\nu\epsilon \end {bmatrix}.
\end{eqnarray}

By projecting the strain tensor onto the irreducible representation of $D_{3d}$, the strain coupling within the framework of linear elasticity theory can be given by
\begin{eqnarray}\label{MA9}
H_{\text{strain}}=\sum_{r}V_{r}\epsilon_{r},
\end{eqnarray}
where $r$ runs over the irreducible representations.
It can be shown that the only contributing representations are the one-dimensional representation $A_{1g}$ and the two-dimensional representation $E_{g}$.
\begin{eqnarray}\label{MA10}
\epsilon_{A_{1g}}&=&t_{\bot}(\epsilon_{xx}+\epsilon_{yy})+t_{\|}\epsilon_{zz} \notag\\
\epsilon_{E_{gx}}&=&d(\epsilon_{xx}-\epsilon_{yy})+f\epsilon_{zx} \notag\\
\epsilon_{E_{gy}}&=&-2d\epsilon_{xy}+f\epsilon_{yz},
\end{eqnarray}
where $t_{\bot}$, $t_{\|}$, $d$, $f$ are four strain-susceptibility parameters. It is valid to drop the $f$ terms due to $f/d\sim10^{-4}$.
The effects of these strain components on the electronic states are described by
\begin{eqnarray}\label{MA11}
V_{A_{1g}}&=&|e_{x}\rangle\langle e_{x}|+|e_{y}\rangle\langle e_{y}| \notag\\
V_{E_{gx}}&=&|e_{x}\rangle\langle e_{x}|-|e_{y}\rangle\langle e_{y}| \notag\\
V_{E_{gy}}&=&|e_{x}\rangle\langle e_{y}|+|e_{y}\rangle\langle e_{x}|.
\end{eqnarray}
Since $V_{A_{1g}}$ is the identity matrix, the $\epsilon_{A_{1g}}$ term only gives rise to the uniform energy shift of all states and can be neglected.
Finally, we rewrite the strain Hamiltonian in the basis of $|e_{+}\rangle$, $|e_{-}\rangle$
\begin{eqnarray}\label{MA12}
H_{\text{strain}}=d(\epsilon_{xx}-\epsilon_{yy})(L_{-}+L_{+}),
\end{eqnarray}
where $L_{+}=L_{-}^{\dag}=|3\rangle\langle1|+|2\rangle\langle4|$ is the orbital raising operator.
We find the common form of the coupling for four kinds of SiVs, and the difference is only the strength of the coupling: $g_{a}=\frac{1}{3}(1+\nu)\epsilon d$ and $g_{t}=(1+\nu)\epsilon d$ corresponding to ($[111]$, $[\bar{1}\bar{1}1]$) and ($[\bar{1}11],[1\bar{1}1]$) orientations.
Finally, by using a rotating wave approximation, we get the Hamiltonian in the text
\begin{eqnarray}\label{MA13}
H_{\text{strain}}=gaJ_{+}+\text{H.c.},
\end{eqnarray}
where $J_{-}=(J_{+})^{\dag}=|1\rangle\langle3|+|2\rangle\langle4|$ is the spin-conserving lowering operator.
For the SiV center in $[\bar{1}11],[1\bar{1}1]$ orientations, the coupling strength is
\begin{eqnarray}\label{MA14}
g=(1+\nu)\epsilon d\sim2\pi\times10\,\text{MHz}.
\end{eqnarray}

\section{Effects of magnetic fields}
To avoid the non-uniform coupling, we adjust the magnetic field to pick up the color centers in a certain orientation.
The static magnetic field $\vec{B}=(B_{x},B_{y},B_{z})$ is represented in the internal coordinate of the four color centers.
In the basis of $|e_{x},\downarrow\rangle,|e_{x},\uparrow\rangle,|e_{y},\downarrow\rangle$ and $|e_{y},\uparrow\rangle$, we neglect the small orbital Zeeman effect and the Hamiltonian of the single SiV center is given by
\begin{widetext}
\begin{eqnarray}\label{MB1}
H_{\text{SiV}}=\begin{bmatrix} \Upsilon_{x}+\frac{1}{2}\gamma_{S}B_{z}&\frac{1}{2}\gamma_{S}B_{x}-\frac{1}{2}i\gamma_{S}B_{y}&\Upsilon_{y}-\frac{1}{2}i\lambda&0 \\ \frac{1}{2}\gamma_{S}B_{x}+\frac{1}{2}i\gamma_{S}B_{y}&\Upsilon_{x}-\frac{1}{2}\gamma_{S}B_{z}&0&\Upsilon_{y}+\frac{1}{2}i\lambda \\ \Upsilon_{y}+\frac{1}{2}i\lambda&0&-\Upsilon_{x}+\frac{1}{2}\gamma_{S}B_{z}&\frac{1}{2}\gamma_{S}B_{x}-\frac{1}{2}i\gamma_{S}B_{y} \\ 0&\Upsilon_{y}-\frac{1}{2}i\lambda&\frac{1}{2}\gamma_{S}B_{x}+\frac{1}{2}i\gamma_{S}B_{y}&-\Upsilon_{x}-\frac{1}{2}\gamma_{S}B_{z} \end {bmatrix}
\end{eqnarray}
\end{widetext}
The eigenvalues of the above matrix can be directly solved as
\begin{widetext}
\begin{eqnarray}\label{MB2}
E=\pm\sqrt{\Upsilon^{2}+\frac{1}{4}\lambda_{SO}^{2}+\frac{1}{4}\gamma_{S}^{2}B_{z}^{2}\pm\sqrt{\gamma_{S}^{2}B_{z}^{2}(\Upsilon^{2}+\frac{1}{4}\lambda_{SO}^{2})+\frac{1}{4}(\gamma_{S}^{2}B_{x}^{2}+\gamma_{S}^{2}B_{y}^{2})^{2}}}
\end{eqnarray}
\end{widetext}
where $\Upsilon=\sqrt{\Upsilon_{x}^{2}+\Upsilon_{y}^{2}}$.
For example, we choose the direction of the magnetic field along the $[1\bar{1}1]$ direction, and the magnitude satisfies $\gamma_{S}|\vec{B}|=2\pi\times20\,\text{GHz}$.
For the SiV in the $[1\bar{1}1]$ direction, the energy level splittings  are
\begin{eqnarray}\label{MB3}
\Delta \omega_{41}/2\pi\approx66\,\text{GHz}
\end{eqnarray}
\begin{eqnarray}\label{MB4}
\Delta \omega_{32}/2\pi\approx26\,\text{GHz}.
\end{eqnarray}
For the other three kinds of SiV centers, these values become
\begin{eqnarray}\label{MB5}
\Delta \omega_{41}^{'}/2\pi\approx53\,\text{GHz}
\end{eqnarray}
\begin{eqnarray}\label{MB6}
\Delta \omega_{32}^{'}/2\pi\approx39\,\text{GHz}.
\end{eqnarray}
If we choose the frequencies of the driving fields as $\omega_{1}/2\pi\approx66$ GHz and $\omega_{2}/2\pi\approx26$ GHz, only the SiV centers in the $[1\bar{1}1]$ direction can be excited, and the other SiV centers are off-resonant with the driving field.
Therefore, Only the SiV centers in the $[1\bar{1}1]$ direction are selected to participate in the Raman process.
In addition, the orbital splitting for all the color centers is $\omega/2\pi\approx46\,\text{GHz}$.

\end{appendix}
%

\end{document}